# Bridging Formal and Perceived Fairness: Development of an Interdisciplinary Framework in Algorithmic Decision-Making

MAIKE LINDERMAYR, Johannes Gutenberg University, Germany
MATTIA CERRATO, Johannes Gutenberg University, Germany
LUISA HÜBNER, Johannes Gutenberg University, Germany
JOHANNES KRAUS, Johannes Gutenberg University, Germany

While fairness has become a central concern in research on algorithmic systems, the field remains predominantly shaped by Computer Science, resulting in a strong emphasis on formal fairness metrics and bias mitigation strategies. Nevertheless, this focus may obscure a fundamental challenge: fairness is not merely a technical property, but a subjective, context-sensitive human judgment shaped by cognitive heuristics, mental models, normative expectations, and sociotechnical factors. Crucially, users' perceptions of fairness may diverge substantially from the fairness criteria an algorithm formally satisfies; a system may meet predefined technical fairness requirements yet still be perceived as unjust by decision-affected stakeholders. In such cases, the system fails on a fundamental dimension: it will not be trusted, accepted, or considered legitimate. Taking a user-centered design perspective, this paper presents a work-in-progress conceptual framework that bridges Computer Science approaches to formal algorithmic fairness with normative and Social Science fairness approaches regarding perceived fairness, trust, and technology acceptance, embedding both within the sociotechnical conditions that shape human judgment. Through (1) theoretical literature synthesis, (2) interdisciplinary workshops, and (3) stakeholder interviews, the project aims to inform evaluation approaches that integrate computational fairness audits with user-centered assessments and guide the design of fairness-aware, human-centered algorithmic systems that support informed, well-calibrated fairness judgments by those affected.

## 1 Introduction and Background

As algorithmic systems increasingly shape high-stakes decisions in domains such as recruitment, finance, and healthcare, algorithmic fairness has emerged as a prominent field of research, predominantly driven by Computer Science [1]. Its growing significance is reflected in recent publication trends: the AI Index [2] reports that the number of fairness and bias-related papers accepted at major AI conferences doubled from 2023 to 2024, underscoring its status

Authors' Contact Information: Maike Lindermayr, Johannes Gutenberg University, Mainz, Germany, malinder@uni-mainz.de; Mattia Cerrato, mcerrato@uni-mainz.de, Johannes Gutenberg University, Mainz, Germany; Luisa Hübner, Johannes Gutenberg University, Mainz, Germany, lhuebn@students.uni-mainz.de; Johannes Kraus, Johannes Gutenberg University, Mainz, Germany, jokraus@uni-mainz.de.

as both an academic focus and a critical R&D priority. The dominant paradigm remains, however, overwhelmingly shaped by Computer Science: formal criteria such as demographic parity [3] and equalized odds [4] define fairness as measurable, mathematical properties of algorithmic outputs – precise, auditable, and essential for regulatory compliance. Yet formal fairness criteria alone are insufficient. Fairness is fundamentally a human judgment, shaped by the values, experiences, and individual cognitive processing of those who design, deploy, and are affected by algorithmic decisions, as well as by organizational constraints, domain-specific norms, and broader societal context [5, 6]. We argue that mental models – commonly defined as individuals' internal representations of what a system contains, how it works, and why it works in a particular way [7] – play a central role in fairness perception. Specifically, we hypothesize that individuals evaluate fairness primarily against their own understanding of how a system operates, which features it relies on, and how its decisions are produced. Dual-process theories of information processing, such as the Elaboration Likelihood Model (ELM) [8], provide a theoretical basis for explaining how these mental models are formed and why they may be incomplete or inaccurate. The ELM distinguishes between a central route, in which judgments result from high elaboration of the available information, and a peripheral route, in which judgments are formed based on superficial cues and simple decision rules [8]. These simple decision rules are known as "heuristics" and reduce the cognitive load required to form a judgment [8, 9, 10]. Applied to fairness assessments, heuristic processing allows individuals to assess the fairness of a system in a cognitively economical manner, without engaging in a full evaluation of a system's properties [9, 10]. The extent of elaboration – thus which route is taken – depends on factors such as motivation, cognitive capacity, and the perceived relevance of the judgment [8]. This perspective implies that fairness-related mental models may vary substantially across users and contexts. Prior work in eliciting and extracting mental models of AI systems has shown that mental models are highly specific to the AI system under consideration. Mental models have been examined across diverse use cases, including augmented decision-making [11], games [12], and fake news detection [13]. Across these contexts, mental models have been used to capture different facets of AI systems, including their reasoning processes, limitations, and outputs [13], error boundaries [11], as well as knowledge distribution and global versus local system behavior [12]. Yet, to our knowledge, research explicitly examining the fairness-related dimensions of users' mental models of AI systems remains scarce. If fairness judgments are shaped by individuals' own, often incomplete or inaccurate, mental models, examining how these models are formed, what shapes them, and where gaps or misconceptions arise can help explain variation in fairness perceptions, including why systems that satisfy formal fairness benchmarks may not be perceived as fair, trustworthy, or acceptable [14, 15, 16].

We therefore propose shifting the discussion beyond formal fairness criteria alone toward the subjective cognitive processes through which fairness properties are perceived and evaluated: the assessment of perceived fairness as a subjective perception of appropriateness that is "downstream from justice" [17, 18]. Drawing on Social Science and justice theories, individuals evaluate fairness along multiple dimensions: distributive justice [19] concerns whether outcomes align with prevailing allocation norms like equity, equality, or need; procedural justice [20] concerns whether the decision process is transparent, consistent, and impartial; while interactional justice [21, 22] concerns whether individuals receive adequate explanations and respectful treatment. These judgments depend not only on formal system properties, but also on how individuals interpret the system and its decision process in light of prior experiences (e.g., with discrimination), institutional trust, or domain-specific expectations [5, 6]. Synthesizing formal fairness measures with normative and Social Science fairness approaches can reveal both conceptual overlaps and systematic divergences. For instance, distributive justice shows a conceptual affinity with demographic parity, as both require that group membership should not produce differential outcomes. However, formal metrics may fail especially regarding the interactional and procedural dimensions, such as respectful treatment, adequate explanation, and the right to contest.

Taken together, both perspectives are necessary and complementary: formal criteria ensure auditability and accountability, while perceived fairness determines whether systems are trusted, accepted, and ultimately used. Only by integrating both can we develop systems that are not merely technically compliant but enable stakeholders to form informed and well-calibrated judgments of their fairness. Yet an integrated framework systematically linking both, while accounting for the sociotechnical conditions under which fairness perceptions emerge, is lacking, and research remains fragmented across Computer Science, Law, Ethics, and Social Science. This paper addresses this gap through a user-centered design lens, which seeks to model users' cognitive processes, values, and contexts as first-class inputs to fairness evaluation. We accordingly argue that a fair algorithmic system must satisfy two conditions simultaneously: formal verifiability across the full algorithmic pipeline, and empirical validation through the subjective fairness assessments of the individuals subject to its decisions.

## 2 Expected Contributions

On the theoretical side, the proposed framework aims to advance theoretical understanding by integrating formal fairness criteria with normative and social science perspectives on fairness perception, including information processing, mental models, and cognitive load. Its core theoretical contribution lies in articulating why systems that satisfy predefined technical fairness requirements do not necessarily translate into perceived fairness: users do not evaluate fairness only through the lens of the chosen formal fairness metric, but through experiential, normative, and relational judgments that formal criteria systematically underspecify. By explicitly mapping the gap between formal and perceived fairness and identifying mediating sociotechnical factors, the framework provides a novel interdisciplinary conceptual structure for cross-disciplinary research questions and hypotheses linking Computer Science, Human Factors, Social Science and Law.

On the practical side, the framework is proposed to support the development of hybrid fairness evaluation methodologies, combining computational audits of formal criteria with structured user-centered perceptual assessments, thereby aiming to reduce the risk of deployment failures arising from perceived unfairness. Concretely, it is expected to inform design guidelines for fairness-aware, human-centered algorithmic systems: guidelines specifying not only which formal properties a system should satisfy, but how decisions should be communicated, contextualized, explained, and made contestable to ensure they are perceived as legitimate by those they affect. In this way, the framework seeks to operationalize a user-centered design approach to algorithmic fairness, one in which the human experience of fairness is a first-order design criterion, not an afterthought.

## 3 Methodological Approach

The framework is developed through three complementary components, outlined below:

(1) First, we conduct a structured interdisciplinary synthesis of the literature on formal fairness criteria, psychological and justice theories, sociotechnical factors, and outcome constructs including trust, acceptance, and behavioural intention. Disciplinary differences are systematically mapped: Computer Science formalizes fairness through mathematical metrics such as demographic parity and equalized odds, while normative and Social Science approaches foreground users' subjective justice perceptions - distributive fairness, procedural fairness, and interactional dimensions such as informational transparency - alongside context-dependent legitimacy judgments. This synthesis identifies conceptual overlaps, tensions, and trade-offs that formal criteria cannot capture, yielding a preliminary construct taxonomy as the framework foundation.

(2) Second, we organize structured workshops with researchers primarily from Computer Science and Human Factors and Engineering Psychology, with other disciplines (e.g., Law, Ethics) invited as needed, to build mutual understanding of each discipline's perspective and collaboratively derive the framework. Workshop sessions pursue three shared goals: establishing common operational definitions of fairness constructs across disciplines; identifying sociotechnical variables that mediate perceived fairness, trust, and acceptance; and resolving disciplinary tensions through negotiated conceptual integration. Participants engage in structured mapping exercises and iterative revision cycles. The anticipated output is a framework specifying structural relationships, encompassing direct effects, mediating variables, and moderating factors between formal fairness properties, sociotechnical mediators, and user-centered perceptual outcomes.

(3) Third, the framework is applied to a selected algorithmic decision-making domain. We conduct qualitative semi-structured interviews with diverse stakeholder groups (e.g., caseworkers, applicants, citizens, or developers). The interviews explore stakeholders' mental models as well as normative attributions regarding which input attributes are perceived as legitimate in algorithmic decisions, how these evaluations differ across stakeholder groups, and how they are shaped by prior experience, transparency, and societal values. Findings will be analyzed to refine the framework, identify unmodeled variables, and derive user-centered design guidelines for fairness-aware algorithmic systems.

## 4 Current Status and Future Directions

This project is currently in its early theoretical phase. Figure 1 outlines the methodological approach and visualizes an initial framework, building on the theoretical foundations detailed before. On the side of the AI system, the framework highlights two interrelated sources of fairness information: fairness-related system properties (e.g., fairness metrics, interface design, explainable AI) and the actors responsible for shaping them, including developers and deploying institutions. The AI system properties thereby are processed in a bottom-up manner: users process information about the system via either a central or a peripheral route, as explicated in the Elaboration Likelihood Model [8], laying the foundation for the establishment of the mental model of the AI-based decision-making system that is further shaped by prior knowledge and assumptions in a top-down manner, e.g., the comparison against a norm set. This comparison results in an overall fairness judgment concerning both the specific decision and the system as a whole, with distributive, procedural, interpersonal, and informational dimensions weighted differently across individuals. The resulting judgment perception is assumed to be shaped by trust, including trust derived from others' fairness judgments of the system as well as trust placed in the developer or institution. The fairness judgment, in turn, is expected to contribute to downstream outcomes such as intention to use and acceptance of the system, without implying that these outcomes are determined by fairness perceptions alone. Rather, we conceptualize the relationship as potentially reciprocal: acceptance and other system-related evaluations may also reinforce or revise fairness judgments over time. Critically, the entire perception-to-judgment process is shaped by sociotechnical, moderating variables – variables affecting the strength of relationships of concepts in the model – such as time pressure, motivation or AI literacy. These moderators affect cognitive processing, for example by influencing the extent to which individuals rely on cognitive heuristics rather than more systematic evaluation. Subsequent phases will pursue quantitative validation through experimental and survey-based studies, testing hypothesized relationships between formal fairness properties, sociotechnical mediators, and fairness perceptions. As a first critical step, we focus on capturing users' fairness-related mental models of a specific AI system and on how to elicit and assess them. We invite conference feedback to critically examine the proposed framework, refine our conceptual approach, and identify promising directions for integrating technical and human-centered perspectives in algorithmic fairness research.

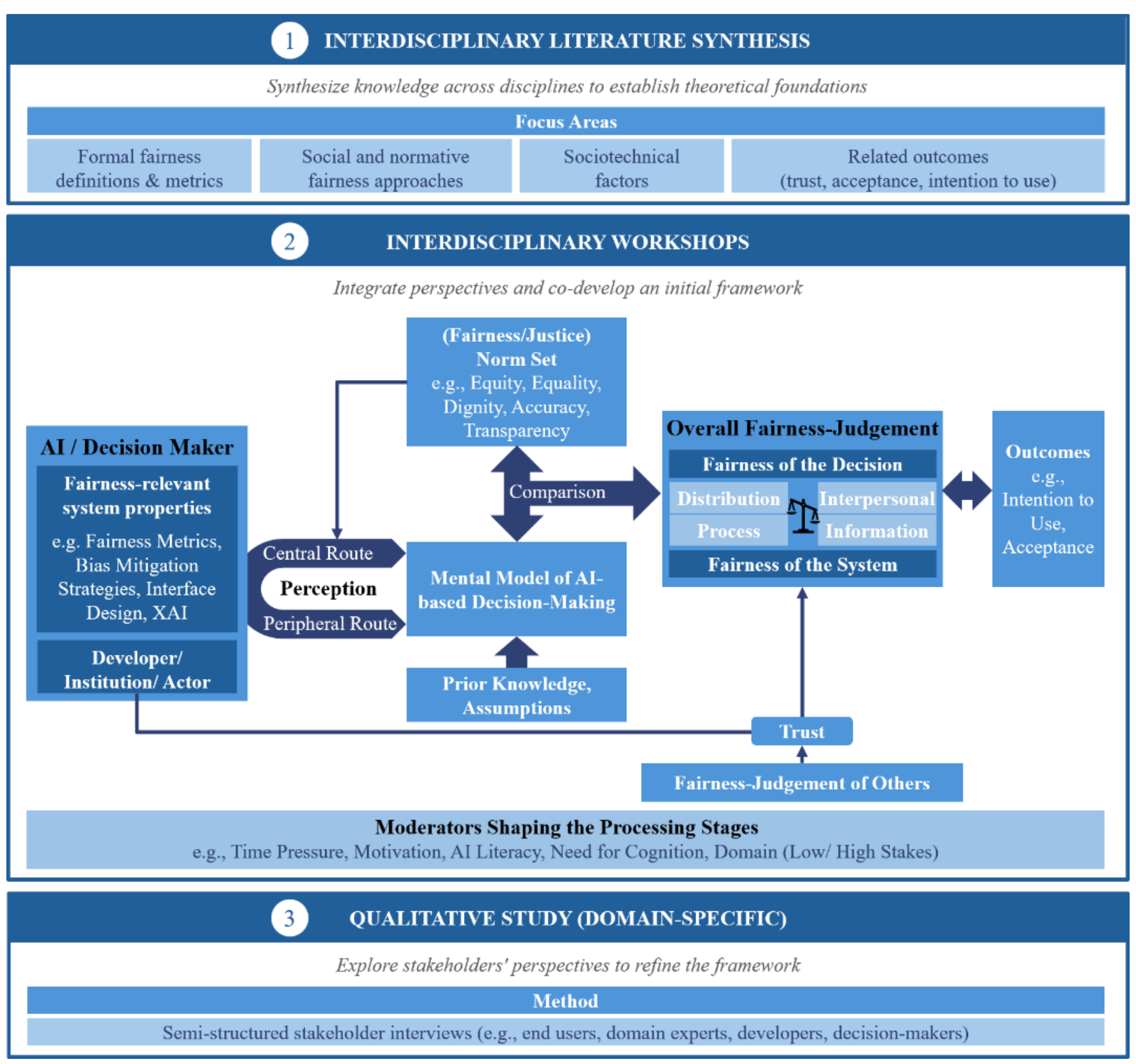

Fig. 1. Methodological approach: Three-phase process for developing an interdisciplinary framework that links formal fairness criteria, sociotechnical factors, and perceived fairness.